\documentclass[twocolumn]{aa}

\pdfoutput=1         

\usepackage{graphicx}
\usepackage{txfonts}

\begin{document}

\title{Testing the evolution of the DB white dwarf GD~358: First results 
of a new approach using asteroseismology}

\author{Jos\'e Miguel Gonz\'alez P\'erez\inst{1,2}
  \and Travis S. Metcalfe\inst{3}}

\offprints{J. M. Gonz\'alez P\'erez, \email{jgperez@iac.es}}

\institute{Instituto de Astrof\'isica de Canarias, 38200 La Laguna, 
Tenerife, Spain 
\and GRANTECAN S.A. (CALP), 38712 Bre\~{n}a Baja, 
La Palma, Spain 
\and High Altitude Observatory, National Center for 
Atmospheric Research, P.O. Box 3000, Boulder, Colorado, USA}

\date{Received (date) / Accepted (date)}

\abstract
{}
{We present a new method that investigates the evolutionary history of the 
pulsating DB white dwarf GD~358 using asteroseismology. This is done 
considering the internal C/O profile, which describes the relative 
abundances of carbon and oxygen from the core of the star to its surface. 
Different evolutionary channels lead to the generation of different C/O 
profiles, and these affect the pulsation periods.}
{We used C/O profiles associated with white dwarfs that evolved through 
binary evolution channels where the progenitor experienced one or two 
episodes of mass loss during one or two common envelope (CE) phases, and 
two profiles from single star evolution. We computed models using these 
different profiles and used a genetic algorithm (GA) to optimize the 
search in the parameter space in order to find the best-fit to the 
observed pulsation periods. We used three-parameter models, adjusting the 
stellar mass ($M_{\star}$), the effective temperature ($T_{eff}$), and the 
helium mass of the external layer ($M_{He}$).}
{Our results suggest that binary evolution profiles may provide a better 
match to the pulsation periods of GD~358. The best-fit to the 
observations is obtained using a profile related to an evolutionary 
history where two episodes of mass loss happen during two CE phases, the 
first during the RGB (Red Giant Branch) stage. The values obtained are 
$T_{eff}$=24~300~K , $M_{\star}$=0.585~$M_{\sun}$ , and 
$\log(M_{He}/M_{\star})$=-5.66. The best-fit model has a mass close to the 
mean mass for DB white dwarfs found in various works, and a temperature 
consistent with UV spectra obtained with the IUE satellite.} 
{}


\titlerunning{Testing the evolution of GD~358}
\authorrunning{Gonz\'alez P\'erez, and Metcalfe}

\maketitle

\section{Introduction}

Pulsating stars are found among white dwarfs. Almost every star in our 
Galaxy will eventually become a white dwarf. Since they are relatively 
simple compared to main sequence stars, white dwarfs provide one of the 
best opportunities for learning about stellar structure and evolution. 
Their interior structure contains a record of the physical processes that 
operate during the later stages in the lives of most stars, so there is a 
potential amount of information encoded in their pulsation frequencies. 
The observational requirements of white dwarfs asteroseismology have been 
addressed by the development of the Whole Earth Telescope (Nather et al. 
1990).

There are presently three known classes of pulsating white dwarfs. The 
hottest class is the planetary nebula nucleus variables, which have 
atmospheres of ionized helium and are also called DOVs. These objects show 
complicated and variable temporal spectra (Gonzalez Perez et al. 
2006), and require detailed calculations that evolve a main sequence 
stellar model to the pre-white dwarf phase to yield accurate pulsation 
periods. The two cooler classes are the helium-atmosphere variable (DBV) 
and hydrogen-atmosphere variable (DAV) white dwarfs. The pulsation periods 
for these objects can be calculated accurately by evolving simpler, less 
detailed models. The DAV stars are generally modeled with a core of carbon 
and oxygen with an overlying blanket of helium covered by a thin layer of 
hydrogen on the surface. The DBV stars are the simplest of all, with no 
detectable hydrogen and only a helium layer surrounding the carbon/oxygen 
core. Robinson et al. (1982) and Kepler (1984) demonstrated that the 
variable white dwarf stars pulsate in non-radial gravity modes.

The chemical evolution of  the DB white dwarfs  atmospheres cannot be 
satisfactorily explained to date (see Shipman 
1997 for a review). In particular, the presence of the DB gap, 
which is the absence of DB white dwarfs between temperatures of 
$\sim$30~000~-~45~000~K (e.g. Liebert 1986), is poorly understood. 
This DB gap persists in the latest results from the Sloan Digital Sky Survey 
(Kleinman et al. 2004). Eisenstein et al. (2006) found several DBs in the gap, 
but still a decrease in the relative number. 
One possibility is that DBs might turn into DAs 
(white dwarfs with pure H atmospheres) through diffusion of small traces 
of H to the surface of the star, concealing the He atmosphere. However, 
Castanheira et al. (2006) showed that atmospheric contamination with H is 
not directly proportional to $T_{eff}$ for DB stars, in conflict with what 
would be expected under the diffusion scenario.

It is also not clear whether DBs are mostly produced by single-star 
evolution or whether a significant fraction of them originate from binary 
progenitors. Nather et al. (1981) pointed out that interacting binary 
white dwarfs (IBWDs) should produce DBs at the end of their evolution. 
AM~CVn, one of the best studied IBWDs, has an $T_{eff} \sim 25 000~K$. 
This temperature is inside the interval where DBVs are found. If AM~CVn 
ends mass transfer phase at this temperature, there will be a single DB 
left, quite possibly a DBV. The examination of these problems can be aided 
by asteroseismology. Compared with other classes of pulsating stars, the 
DBVs are promising candidates for asteroseismology: their mode spectra are 
relatively rich, and pulsation theory for these stars is quite advanced 
and well tested.

GD~358, also called V777~Herculis, is the prototype of the DBVs and is one 
of the classical examples of the application of asteroseismological 
methods (e.g. Winget et al. 1994; Vuille et al. 2000; Kepler et al. 2003). 
GD~358 was the first pulsating star detected based on a theoretical prediction 
(Winget et al. 1985). When the Whole Earth Telescope observed GD~358 in 
1990, 181 periodicities were detected, but only modes 
from radial order $k=8$ to 18 were identified, most of them showing 
triplets, consistent with the degree $l=1$ identification (Winget et al. 
1994).  Detecting as many modes as possible 
is important, as each independent mode detected yields an independent 
constraint on the star's structure. The authors derived the total stellar mass 
as $0.61 \pm 0.03$ $M_{\sun}$, the mass of the outer helium envelope as 
$(2.0 \pm 1.0) \times 10^{-6}$ $M_{\star}$, the luminosity as $0.050 \pm 
0.012~L_{\sun}$ and, deriving a temperature and bolometric correction, the 
distance as $42 \pm 3$ pc.

Asteroseismological analysis of GD~358 has been a successful tool to test 
pulsation theory. For instance, the central oxygen abundance of the models 
has been shown to have a measurable effect on the pulsation frequencies 
(Metcalfe et al. 2001 -hereafter M1-), and also the possible presence of a 
$^{3}$He/$^{4}$He transition zone caused by chemical diffusion (Montgomery 
et al. 2001), although more recent observations have ruled out this 
possibility (Wolff et al. 2002).

\section{Pulsations and C/O profiles.}

The interior structure of the white dwarfs contains a record of the 
physical processes that operate during the later stages in the lives of 
most stars. The distribution of the C/O ratio depends on the evolution of 
the star. Stellar evolution theory predicts that a large number of white 
dwarfs are post Asymptotic Giant Brach (AGB) stars. Therefore, they should 
consist of the primary ashes of He-burning, i.e. carbon and oxygen. The 
central ratio C/O depends on the interplay between the triple-$\alpha$ and 
the $^{12}C(\alpha,\gamma)^{16}O$ reactions during the core helium burning 
phase. In the first part of the burning the most efficient process is the 
triple-$\alpha$, which produces $^{12}$C. When the central abundance of 
helium decreases below about 0.15~$M_{\star}$, the production of $^{16}$O 
via the $^{12}C(\alpha,\gamma)^{16}O$ becomes the dominant process.

While they are still embedded in the cores of red giant models, the 
internal chemical profiles of white dwarfs models show a relatively 
constant C/O ratio near the center, with a size determined by the extent 
of the helium-burning convective region. Further out the oxygen mass 
fraction decreases as a result of the helium-burning shell moving towards 
the surface of the red giant model while gravitational contraction causes 
the temperature and density to rise. This increases the efficiency of the 
triple-$\alpha$ reaction, producing more carbon relative to oxygen. The 
central oxygen mass fraction is lower in higher mass white dwarfs models. 
The rate of the triple-$\alpha$ reaction increases faster at higher 
densities than does the $^{12}C(\alpha,\gamma)^{16}O$ reaction. As a 
consequence, more helium is used up in the production of carbon, and 
relatively less is available to produce oxygen in higher mass models.

Metcalfe et al. (2000 -hereafter M0-) used the observed periods of GD~358 
from Winget et al. (1994) to search for the optimum theoretical model with 
static diffusion envelopes. This search was based in the application of a 
genetic-algorithm-based optimization (GA) to white dwarf pulsation models, 
which performs a global search to provide objective global best-fit models 
for the observed pulsation frequencies. They used a three-parameter model 
including $M_{\star}$, $T_{eff}$, and $M_{He}$, and six different 
combinations of core composition and C/O transition profiles to search for 
the best-fit model. Their results showed that both the central oxygen 
abundance and the shape of the C/O profile affect the pulsation pattern.

M1 modified the code to include as free parameters any central oxygen mass 
fraction ($X_{0}$) between 0 and 1 with resolution 1\% and a fractional 
mass parameter ($q$). The GA fitting process explored different chemical 
profiles built in the following way: the value $X_{0}$ was fixed to its 
central value out to $q$ that varied between 0.1 and 0.85 with resolution 
0.75\%. From this point $X_{0}$ decreased linearly in mass to zero oxygen 
at the 95\% mass point. The results showed that significant improvement to 
the fit was possible by including $X_{0}$ and $q$ as free parameters, 
confirming that the observed pulsations really contain information about 
the hidden interiors of the stars. The best fit obtained was: 
$T_{eff}=22\,600 K$; ~$M_{\star}=0.650 M_{\sun}$; 
~$\log(M_{He}/M_{\star})=-2.74$; ~$X_{0}=0.84$; ~$q=0.49$.

Therefore, the C/O profiles may give us an opportunity to explore the 
previous evolution of the DBV white dwarfs. C/O profiles obtained from 
single evolutionary models are different than those obtained from 
evolutionary models of close binary systems (e.g. Iben \& Tutukov 1985 
-hereafter IT85-; Iben 1986). The shapes of the C/O profiles may be 
affected due to interaction with close companions, i.e., if CE phases 
happen during the RGB and/or AGB stages.

\section{Close binary evolution.}

Peculiar white dwarf stars may be created through binary evolutionary 
channels. We have concentrated on evolutionary channels that produce white 
dwarfs without hydrogen. Hydrogen may be removed of the system due to 
previous CE stages. The resulting white dwarfs may have a surface rich in 
helium and may contribute to the DB population.

IT85 analyzed the possibility to obtain such white dwarfs by evolving 
components of close binary systems where the component masses were in the 
range 2 to 12 $M_{\sun}$. They assumed models where the more massive 
component filled its Roche lobe during the RGB, after the central hydrogen 
exhaustion, but before or near the onset of core helium burning. The 
binary components experience one or two mass loss processes due to CE 
stages, depending upon the initial main-sequence mass.

Components with masses in the range 2.3 to 4.8 $M_{\sun}$ experience one 
episode of mass loss and become degenerate dwarfs of mass in the range 
0.32 to about 0.7 $M_{\sun}$. The remnants experience several small 
hydrogen shell flashes and one large final flash which, for the more 
massive degenerates dwarfs, remove almost all of the hydrogen-rich layer. 
In these cases the total mass of the hydrogen layer is reduced to about 
$10^{-5} M_{\sun}$. This evolutionary channel may produce remnants with 
masses as small as 0.4~$M_{\sun}$ and can convert 70\% of the initial 
helium into carbon/oxygen. This represent a substantial difference from 
evolutionary channels of single low-mass stars, which do not ignite and 
burn helium until the mass of the electron-degenerate cores reaches 
$\sim$0.5 $M_{\sun}$. Han et al.(2000) have performed close binary 
evolutionary calculations and found that a carbon/oxygen white dwarf with 
mass as low as 0.33~$M_{\sun}$ may be formed from stable Roche overflow if 
the initial mass of the primary is close to 2.5~$M_{\sun}$.

IT85 found that binary components of initial mass in the range 4.8 to 10.3 
$M_{\sun}$ lose the hydrogen-rich matter in two episodes of Roche-lobe 
overflow and become C/O degenerate dwarfs with masses in the range 0.7 to 
1.08 $M_{\sun}$. This channel leads to the production of DB white dwarfs. 
For the more massive remnants (over 0.8 $M_{\sun}$), a substantial part of 
the helium layer is also removed from the system during the second 
Roche-lobe overflow, so the degenerate dwarf would have a thin helium 
outer layer. This layer is expected to be thicker for remnants below 0.8 
$M_{\sun}$ . IT85 estimated a formation rate of $\sim10^{-2} yr^{-1}$ of 
non-DA degenerate dwarfs created for these binary scenarios and concluded 
that they may produce 15\%-20\% of all hot non-DA white dwarfs.

\section{Computation of models.}

\subsection{DB white dwarfs models.}

To find the theoretical model of a white dwarf, we start with a static 
model of a pre-white dwarf and allow it to evolve quasi-statically until 
it reaches the desired temperature. We then calculate the adiabatic 
non-radial oscillation frequencies for the output model. It is not 
necessary to go through detailed calculations that evolve a main sequence 
star all the way to its pre-white dwarf phase. For the DB and DA white 
dwarfs, it is sufficient to start with a hot polytrope of order 2/3 (i.e. 
$P~\propto~\rho^{5/3}$). The cooling tracks of these polytropes converge 
with those of the pre-white dwarfs models above the temperatures at which 
DB and DA white dwarfs are observed to be pulsationally unstable (Wood 
1990).

We have used the White Dwarf Evolution Code (WDEC) to evolve a start
model to a specific temperature. This code is described in detail by Lamb 
\& Van Horn (1975) and by Wood (1990). It was originally written by Martin 
Schwarzschild, and has subsequently been updated and modified by many 
others including: Kutter \& Savedoff (1969), Lamb \& Van Horn (1975), 
Winget (1981), Kawaler (1986), Wood (1990), Bradley (1993), and Montgomery 
(1998). The pulsation frequencies of the output models are determined 
using the adiabatic non-radial oscillation (ANRO) code described by 
Kawaler (1986). Practical modifications to these programs were done by M0, 
primarily to allow models to be calculated without any intervention by the 
user.

The equation of state for H/He comes from Saumon et al. (1995). We use the 
OPAL opacity tables from Iglesias \& Rogers (1996), neutrino rates from 
Itoh et al. (1996), and the ML2 mixing-length prescription of B\"{o}hm \& 
Cassinelli (1971). We have fixed the ratio of mixing-length/pressure scale 
height to 1.25, as recommended by Beauchamp et al. (1995). The evolution 
calculations for the core are fully self-consistent, but the envelope is 
treated separately. The core and envelope are stitched together and the 
envelope is adjusted to match the boundary conditions at the interface. 
Adjusting the helium layer mass involves stitching an envelope with the 
desired thickness onto the core before starting the evolution. Because 
this is done while the model is still very hot, there is plenty of time to 
reach equilibrium before the model approaches the final temperature.

\subsection{The genetic algorithm and Darwin.}

The global search for the optimal model parameters to fit the independent 
pulsations periods of GD~358 is done using the GA optimization method 
developed by M0 and M1. This method improves the objective search in the 
parameter space over traditional procedures like iterative methods 
starting from a first guess. This standard approach has a potential 
problem: the initial set of parameters is usually determined subjectively. 
This, combined with a local search in the parameter space, may produce a 
local good fit that is not the global best fit. Restrictions by the GA 
method regarding the range of the parameters space are imposed only by 
observational constraints and the physics of the model. Therefore, GA 
provides a relatively efficient way of searching globally for the best-fit 
model.

The first step of the GA is to fill up the parameter space with trials 
consisting of randomly chosen values for each parameter, within a range 
based on the physics that the parameters are supposed to describe. The 
model is evaluated for each trial, and the result is compared with the 
observational data. This comparison assigns a fitness to the trial 
inversely proportional to the variance. The new generation of trials is 
then created taking into account the previous fitness. The fittest models 
survive to the next generation and the parameter space is then sampled 
more around these values. Operators emulating reproduction and mutation 
are applied to produce the new generation of trials. The evolution 
continues for a specified number of generations, chosen to maximize the 
efficiency of the method. We used a population size of 128 trials and 
allowed the GA to run for 200 generations. We have performed a total of 10 
GA runs for each core composition to reduce the chances of not finding the 
best answer to less than about 3 in 10~000.

The GA code has been implemented in a network of 64 PCs running Linux 
(Metcalfe \& Nather 1999 and 2000). The white dwarf code runs on this 
metacomputer by using the Parallel Virtual Machine (PVM) software (Geist 
et al. 1994), which allows a collection of networked computers to 
cooperate on a problem as if they were a single multiprocessor parallel 
machine. The central computer in the network is called Darwin and runs the 
parallel version of the genetic algorithm (PIKAIA) which is the master 
program. After creating a new generation, Darwin distributes to the slave 
computers an array with the models to check. The slave computers perform 
the following actions: evolve a white dwarf model to the specified 
temperature, determine the pulsation periods of the model, and compare 
observations with the calculated pulsation periods. Then it sends the 
result back to Darwin which will include it in the GA code. The slave 
computer is then ready for a new trial so Darwin can send a new trial from 
the generation.

\subsection{Parameter space.}

We use a three-parameter model including $M_{\star}$, $T_{eff}$, and 
$M_{He}$. The GA search space is defined as follows: the masses are 
confined between 0.45~$M_{\sun}$ and 0.95~$M_{\sun}$. Although 
Kepler et al. (2007) reported a substantial number of white dwarfs over this upper 
limit, Beauchamp et al. (1999) and Castanheira et al. (2006) 
found that all known DBVs appear to have masses within this range. 
The temperature search includes values 
between 20~000~K and 30~000~K and is based in the temperature 
determination for the known DB instability strip, that is, the strip in 
the $T_{eff}$ versus $\log (g)$ diagram where the DB white dwarfs are 
pulsationally unstable (Beauchamp et al. 1999). These authors, depending 
on various assumptions, place the red edge as low as 21~800 K, and the 
blue edge as high as 27~800 K. Finally, the search interval for the mass 
of the atmospheric helium layer has a lower limit $10^{-8}~M_{\star}$ and 
an upper limit $10^{-2}~M_{\star}$. Masses must not be greater than this 
because the pressure on the overlying material will then theoretically 
initiate helium burning at the base of the envelope. At the other extreme, 
none of the models pulsate for helium masses less than $10^{-8}~M_{\star}$ 
over the entire temperature range we consider (Bradley \& Winget 1994).

A two-digit decimal is used for encoding the different parameters. This 
results in a temperature resolution of 100~K, a mass resolution of 
0.005~M$_{\sun}$, and a resolution for the helium layer thickness of 0.05 
dex.

\subsection{Computing profiles.}

\begin{figure}
\resizebox{\hsize}{!}{\includegraphics{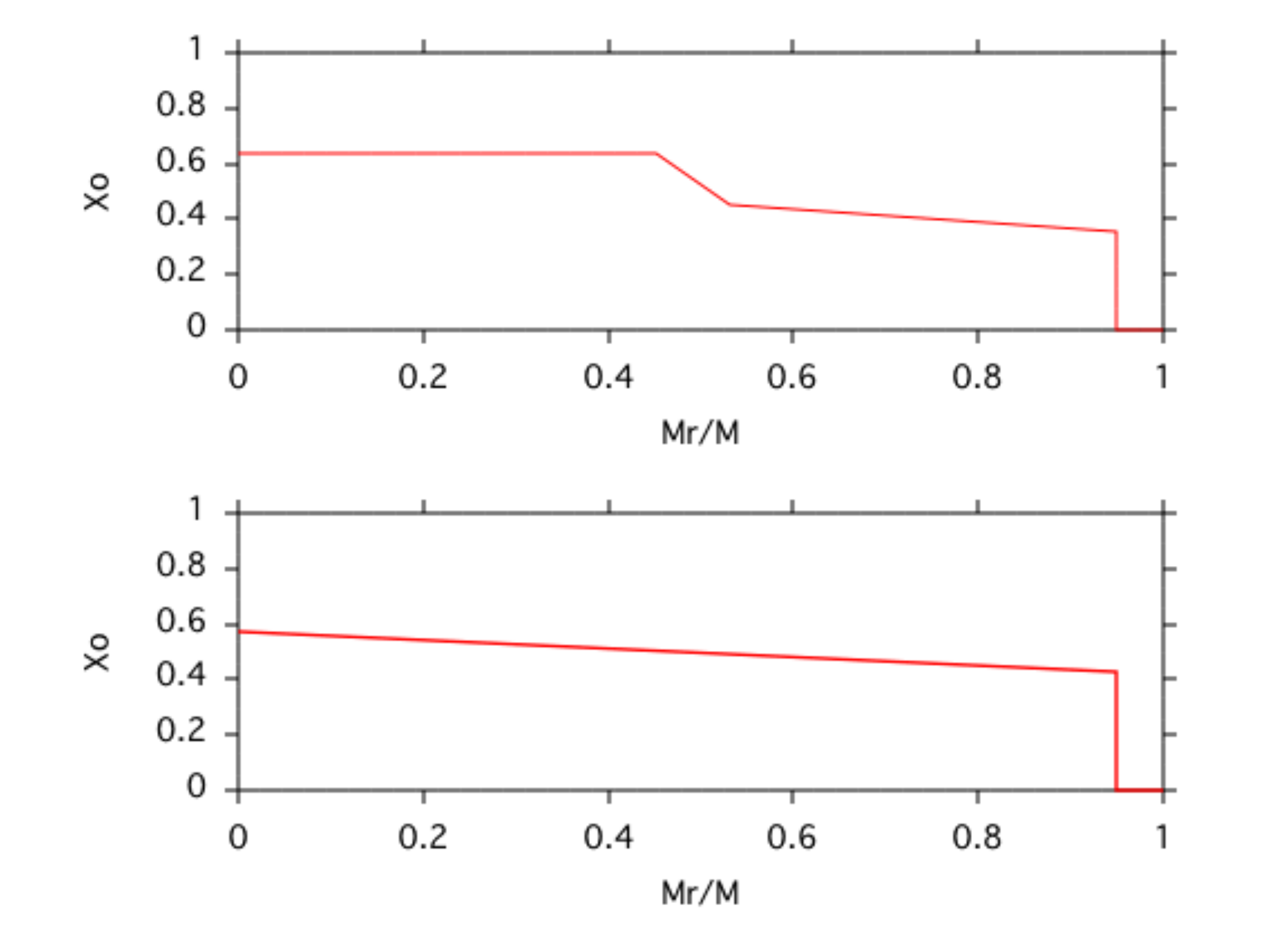}}
 \caption{Profiles of evolutionary channels affected by 2 epochs of mass 
loss during 2 CE phases, the first one during the RGB stage. 
The upper panel shows the profile obtained for a 0.752~$M_{\sun}$ remnant 
and the lower for a 0.894~$M_{\sun}$ remnant.}
 \label{pr_2CE}
\end{figure}
 
\begin{figure}
 \resizebox{\hsize}{!}{\includegraphics{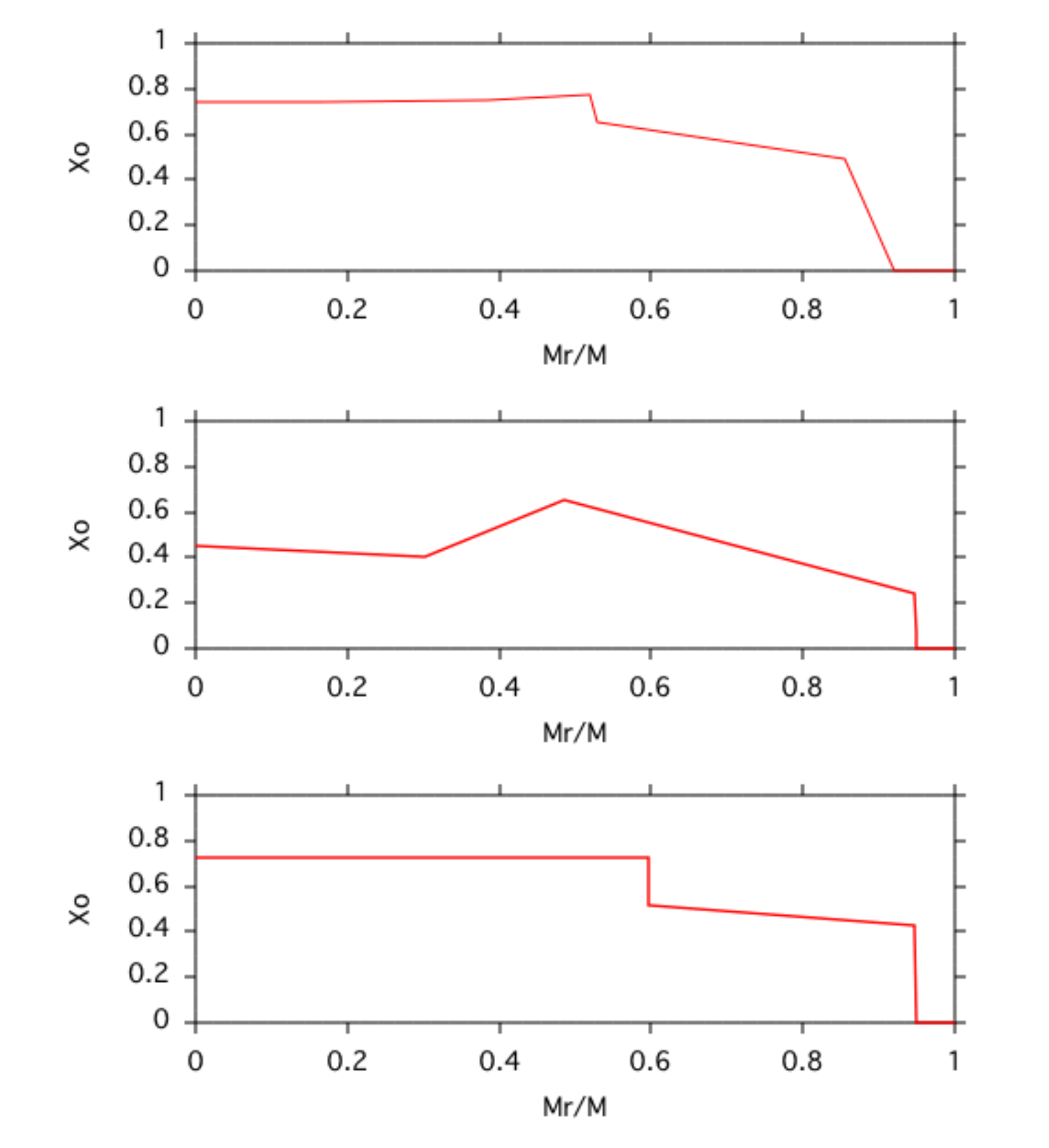}}
 \caption{Profiles of an evolutionary channel affected by 1 CE, and 2 
evolutionary channels of single stars. The upper panel shows the profile 
obtained with a evolutionary history affected by 1 process of mass loss 
due to 1 CE during the RGB stage for a 0.523~$M_{\sun}$ remnant. The 
middle panel shows the profile obtained, using a single evolutionary model, 
for a white dwarf model of mass 0.599~$M_{\sun}$ that evolved from a 
horizontal-branch star of initial mass 7~$M_{\sun}$. The lower panel shows 
another profile related to a single evolutionary scenario: a white dwarf 
model of mass 0.965~$M_{\sun}$ that evolved from a single AGB model star 
of mass 7~$M_{\sun}$}
 \label{pr_1CEandSE}
\end{figure} 

We have used the C/O profiles presented in IT85 and Iben (1986). We had to 
modify the original profiles to be able to incorporate them into our code. 
The axes have been transformed to $X_{O}$ (mass fraction of oxygen) versus 
$M_{r}/M_{\star}$ (enclosed mass fraction at a given radius). A C/O 
profile is computed using a series of pairs of points ($X_{C}$, 
$M_{r}/M_{\star}$), where $X_{C}=1-X_{O}$. These pairs define the profile 
shape in the model. We have used between 6 and 8 points. This procedure 
simplifies the original shape of the profile but reproduces it well. One 
limitation for computing the profiles is that $X_{O}$ must be set to zero 
outside the mass fraction ($M_{r}/M$)=0.95. However, some of the original 
profiles present some amount of oxygen outside this fractional mass. 
Another simplification is that our models consider carbon and oxygen as 
the only nuclear species in the interiors of the white dwarfs. IT85 
evolutionary models present traces of other elements like $^{22}Ne$ and 
$^{25}Mg$. Figures~\ref{pr_2CE} and \ref{pr_1CEandSE} show the profiles we 
tested.

Our study is mainly based on IT85 because, until now, it is the only work 
presenting several C/O profiles for different evolutionary channels of 
close binary systems. In addition, the profiles are obtained using the 
same physics, which improve the possibilities for a quantitative 
comparison. Other profiles for single evolutionary scenarios are 
available (e.g. Salaris et al. 1997). However, the physics included in 
these evolutionary models are slightly different, e.g. how overshooting is 
treated, nuclear reactions rates, etc. Models with different physics may 
produce different C/O profile shapes.

IT85 evolutionary models start with an initial pair of stars with certain 
masses and separation and are evolved to produce a white dwarf with a 
certain C/O profile and mass, mainly without hydrogen on the surface. 
Therefore, each C/O profile tested is already related to a certain mass. 
In our code, $M_{\star}$ is a free parameter. A self-consistent result for 
a given profile would require a best-fit model with a value for 
$M_{\star}$ not too far from the final masses in IT85, though profiles for 
different masses should be qualitatively similar.

\begin{table*} 
\centering 
\caption[]{Characteristics of the evolutionary channels of the profiles 
we tested. The second column indicates how many episodes of mass loss have 
happened due to CE phases in the previous evolution. RGB indicates that 
the first happened during the RGB stage. SE indicates single evolution. HB 
indicates that the white dwarf evolved from a Horizontal Branch model, and 
AGB from an Asymptotic Giant Branch model.}
  \begin{tabular}{llll}
\hline\hline\\[-6pt]
Profile id & Evolution & Mass WD ($M_{\sun}$) & Hydrogen\\[2pt] 
\hline\\[-6pt]
1 & 2 CE, RGB & 0.752 & no\\
2 & 2 CE, RGB & 0.894 & no\\
3 & 1 CE, RGB & 0.523 & $\sim2.6\times10^{-4}$\\
4 & SE, HB & 0.599 & yes\\
5 & SE, AGB & 0.965 & yes\\
\hline
\label{Profiles}
\vspace{0 cm}

\end{tabular}
\end{table*}

Table~\ref{Profiles} presents the properties of the different C/O profiles 
we tested. The second column indicates how many episodes of mass loss have 
happened due to CE phases in the previous evolution. RGB indicates that 
the first CE happened during the Red Giant Brach stage. SE indicates 
single evolution. HB indicates a white dwarf remnant coming from a 
horizontal branch model, and AGB from an Asymptotic Giant Branch model. We 
have included two profiles associated with a degenerate white dwarf that 
has experienced 2 epochs of mass loss in CE episodes, the first one during 
the RGB phase: profile 1 is from the 0.752~$M_{\sun}$ remnant of a model 
of initial mass 5~$M_{\sun}$ (see upper panel of Figure~\ref{pr_2CE}), 
while profile 2 comes from a 0.894~$M_{\sun}$ remnant of a model with 
initial mass 6.95~$M_{\sun}$ (see lower panel of Fig.~\ref{pr_2CE}). 
Profile 3 corresponds to the 0.523~$M_{\sun}$ remnant of a model of 
initial mass 4~$M_{\sun}$ that has experienced only one episode of mass 
loss due to a CE during the RGB phase. This model retains some hydrogen in 
the surface layer, $\sim$$2.6\times10^{-4}$~$M_{\star}$ (see profile in 
the upper panel of Fig.~\ref{pr_1CEandSE}).

We have completed our comparative analysis by checking the best-fit models 
for 2 profiles obtained from single evolutionary scenarios. Profile 4 is 
related to a white dwarf model of mass 0.599~$M_{\sun}$ that comes from 
the evolution of a horizontal-branch model of initial mass 7~$M_{\sun}$ 
(Iben 1982). This is presented in the middle panel of 
Fig.~\ref{pr_1CEandSE}). Profile 5 is the remnant 0.965~$M_{\sun}$ core of 
a single AGB model star of mass 7~$M_{\sun}$ (Iben 1976) (see lower panel 
of Fig.~\ref{pr_1CEandSE}). This last model presents some inconsistency 
since our parameter space for the mass is limited to masses below 
0.95~$M_{\sun}$.

\section{Results.}

\begin{table*}
\centering
\caption[]{Best models obtained for the different evolutionary channels we
tested. The fits labeled with '$a$' are not compatible with the observed 
parallax of GD~358, which constrains the luminosity.}
\begin{tabular}{lllll}
\hline\hline\\[-6pt]
Profile id & $T_{eff}$ (K) & $M/M_{\sun}$ & $\log(M_{He}/M_{\star})$ & 
r.m.s. (s)\\[2pt]
\hline\\[-6pt]
1 & 25~000 & 0.555 & -5.60 & 2.13\\
2 & 24~300 & 0.585 & -5.66 & 2.12\\
3 & 23~700 & 0.605 & -6.03 & 2.19\\
4 & 25~100 & 0.825 & -3.96 & 2.23$^{a}$\\
5 & 26~000 & 0.560 & -5.39 & 2.21\\
\hline
\label{results}
\vspace{0 cm}
\end{tabular}
\end{table*}

\begin{figure*}
\centering
 \includegraphics[width=17cm]{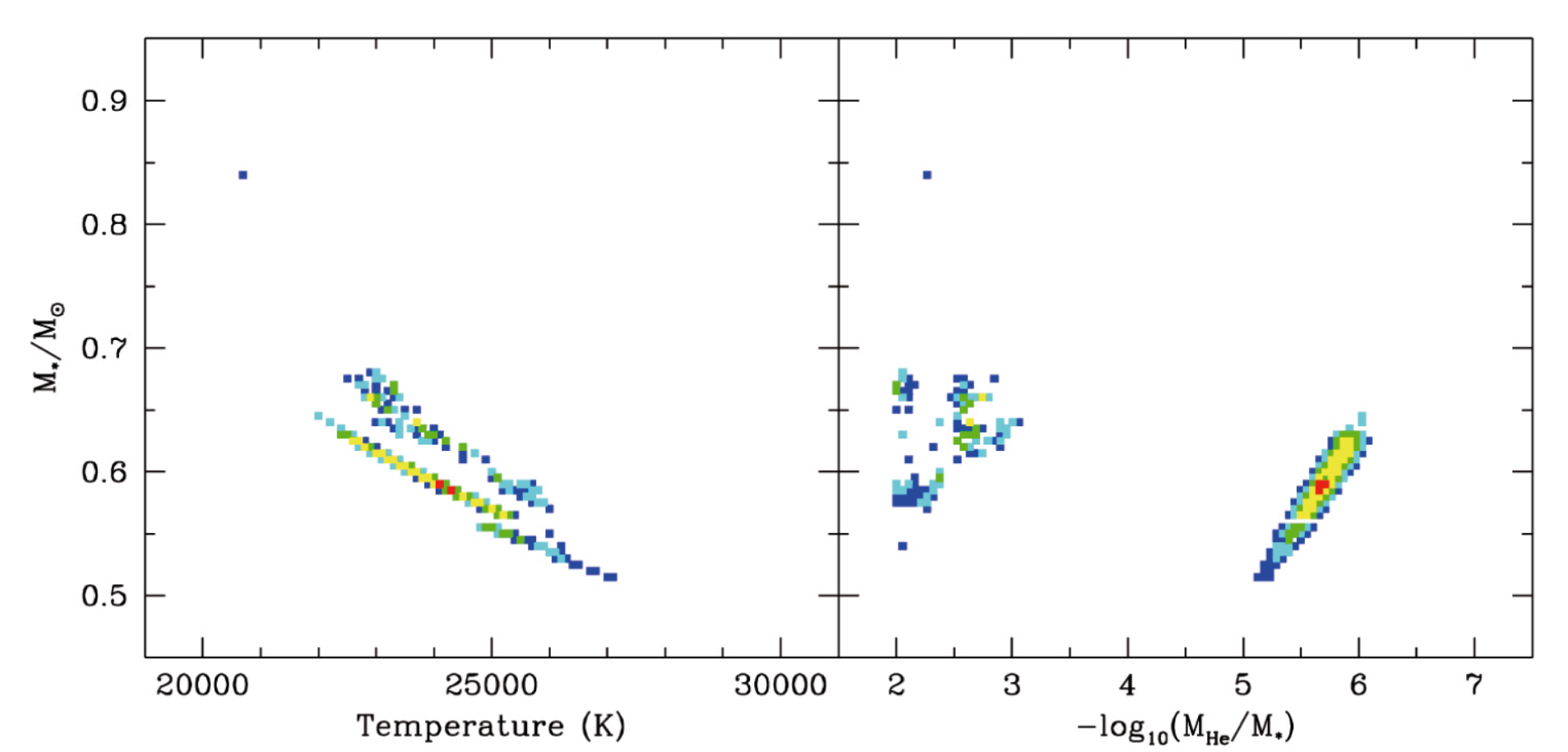}
 \caption{Front and side views of the search space for the best model 
(profile 2). Square points mark the location of models that yields a 
reasonable match to the periods observed for GD~358. The color of a point 
indicates the relative quality of the match (see text for details).}
 \label{best}
\end{figure*}

Table~\ref{results} presents the best fits obtained for each profile. The 
best fits are obtained for the profiles 1 and 2, which involve 
evolutionary channels where the star experiences two CE phases. Profile 1 
is associated in the IT85 evolutionary model to a white dwarf with mass 
0.752~$M_{\sun}$, and profile 2 to a more massive white dwarf (0.894 
$M_{\sun}$). The asteroseismological masses obtained are in both cases 
smaller: 0.555 and 0.585~$M_{\sun}$. Both fits have almost the same 
residuals, 2.13 and 2.12~s, but the values obtained for the 3 parameters, 
$T_{eff}$, $M_{\star}$, and $\log(M_{He}/M_{\star})$, are slightly 
different. This confirms the dependence of the models on the intrinsic 
shape of the profile.

The best fit is obtained using profile 2. The parameters obtained are 
$T_{eff}$=24~300~K , $M_{\star}$=0.585~$M_{\sun}$ , and 
$\log(M_{He}/M_{\star})$=-5.66. The best-fit model has a mass close to the 
mean mass for DB white dwarfs, and a temperature consistent with UV 
spectra obtained with the IUE satellite (Castanheira et al. 2005).

Figure~\ref{best} shows the general characteristics of the 
three-dimensional parameter space: $M_{\star}$, $T_{eff}$, and 
$\log(M_{He}/M_{\star})$. All the combinations of parameters found by the 
GA algorithm for our best fit model with residuals smaller than 3.3~s are 
presented. The two panels are orthogonal projections of the search space, 
so each point in the left panel corresponds to a point in the right panel. 
The fit is displayed using the following color scale: a red point if the 
residuals (r.m.s.) are smaller than 2.2 s; yellow: $2.2 \le r.m.s. < 2.4$; 
green: $2.4 \le r.m.s. < 2.7$; cyan: $2.7 \le r.m.s.<3.0$; blue: $3.0 \le 
r.m.s.<3.3$. The figure shows the presence of more than one region that 
yields a good match to the observations. Two families of good fits, for 
thick and thin helium layers, are displayed, but the amount and quality of 
the fits found for thin layers are larger. The distribution of the best 
models form stripes in the search spaces panels. This is due to the 
parameter correlations in both projections. The correlation between 
$M_{\star}$ and $\log(M_{He}/M_{\star})$ is described by Brassard et al. 
(1992), who showed that a thinner helium layer can compensate for an 
overestimate of the mass.

The correlation between $M_{\star}$ and $T_{eff}$ is related to the 
Brunt-V\"{a}is\"{a}l\"{a} frequency, which reflects the difference between 
the actual an the adiabatic density gradients. The pulsation periods of a 
white dwarf model reflect the average of the Brunt-V\"{a}is\"{a}l\"{a} 
frequency through the star. If the temperature decreases, the matter 
becomes more degenerate, so the Brunt-V\"{a}is\"{a}l\"{a} frequency tends 
to be zero in much of the star, leading to lower pulsation frequencies. So 
an overestimate of the mass can compensate for an underestimate of the 
temperature.

\begin{figure}
\resizebox{\hsize}{!}{\includegraphics{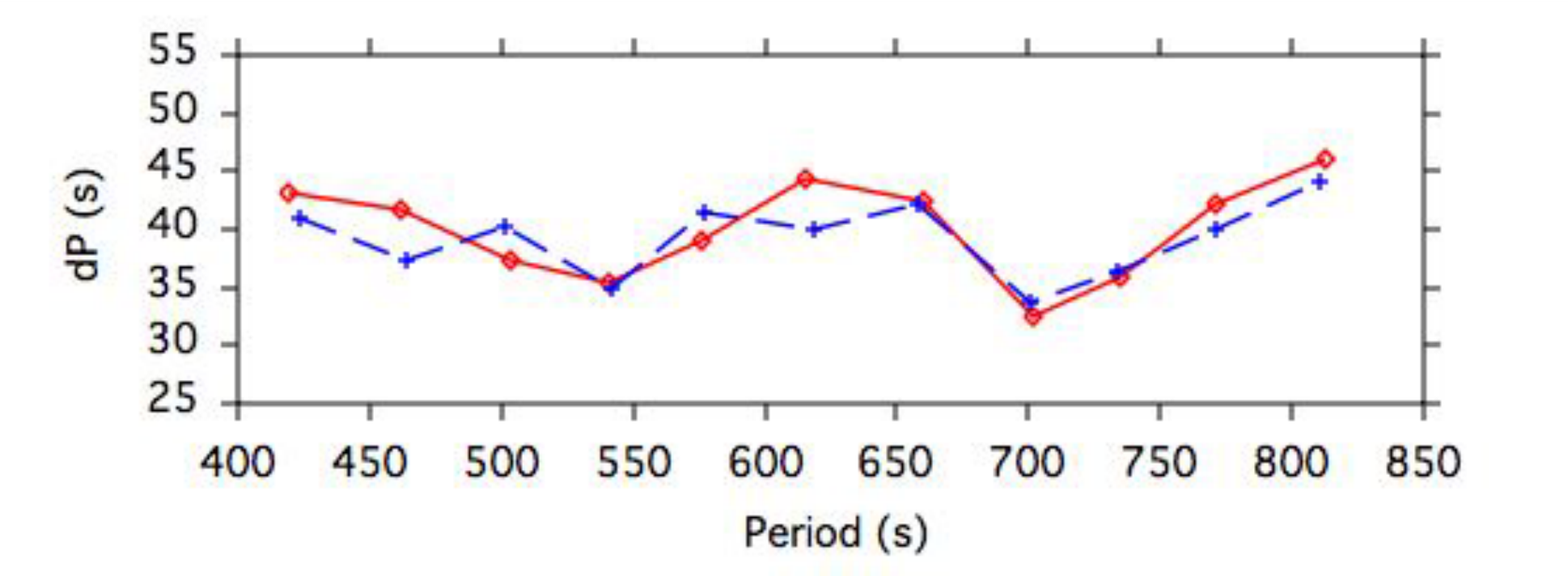}}
 \caption{The observed periods of GD~358 (dashed line), and the optimal 
model periods found by the genetic algorithm (solid line), using a profile 
for a remnant of 0.894 $M_{\sun}$ from binary evolution with 2 CEs.}
 \label{p_dp_CEs}
\end{figure}
 
Figure \ref{p_dp_CEs} presents the observed and calculated periods for the 
best model plotted against the forward period spacing 
($\delta P \equiv P_{k+1}-P_k$). Note that the GA only fits the periods of 
the pulsations modes, and the agreement between the deviations from the 
mean period spacing is a reflection of the overall quality of the match.

\section{Discussion.}

We have presented a new method to investigate the evolutionary history of 
GD~358, which uses the dependence of the pulsation frequencies on the 
internal C/O profile. Different evolutionary channels lead to the 
generation of different profiles describing how the C/O ratio changes from 
the core of the star to its surface. We have tested and compared profiles 
associated to binary evolution where the object experiences one or two 
episodes of mass loss during one or two CE phases. In all the cases the 
first CE happens during the RGB phase. We have also analyzed two profiles 
obtained from single evolution channels for comparison.

Our results confirm the dependence of the pulsations on the shape of the 
C/O profile and indicate that binary evolution may provide a better match 
to the pulsation frequencies of GD~358.  Our best-fit model involves a 
previous evolutionary history with two CE episodes that has removed all of 
the hydrogen from the surface of the star. This is an alternative way to 
explain the lack of hydrogen in DB white dwarfs. IT85 estimated a 
formation rate of $\sim10^{-2} yr^{-1}$ of non-DA degenerate dwarfs 
created from these binary scenarios, and concluded that they may produce 
15\%-20\% of all non-DA white dwarfs. Our results suggests that GD~358 may 
be one of these. The best-fit model has a mass close to the mean mass for 
DB white dwarfs, a temperature consistent with recent calibration of UV 
spectra obtained with the IUE satellite, and a thin outer layer of helium.

The value obtained for $M_{\star}$ in our best-fit model 
(0.585~$M_{\sun}$) is lower than the IT85 mass value related to the 
profile that produced this best-fit (0.894 $M_{\sun}$). In general, there 
are differences between these values for the different profiles tested. 
This may indicate that the physics of the evolutionary code and the white 
dwarf models are not completely consistent. However, in most of the cases, 
we have included profiles that use the same evolutionary code to be able 
to do qualitative comparisons, avoiding possible differences in the fits 
due to the different physics applied to the models. Therefore, the method 
presented may also be used to investigate the coherence between 
evolutionary codes and the adiabatic non-radial oscillation code. Further 
C/O profiles obtained with improved physics must be analyzed to 
investigate this relation.

The results obtained for GD~358 indicate that this method may be a good 
option to investigate the evolution of DBV white dwarfs. This method is 
strongly related to observational properties of the stars, in particular 
to their pulsation periods. The main problem is that it can only be 
applied to few DBV white dwarfs: those that have been extensively observed 
in multisite campaigns in order to resolve their temporal spectrum. 
Currently there are only two candidates suitable for this study: GD~358 
and CBS~114 (Handler et al. 2002; Metcalfe et al. 2005). Further options 
for the future include extending this study using profiles coming from 
double evolutionary scenarios where the star experienced an epoch of mass 
loss from one CE episode during the AGB phase.


\begin{thebibliography}{}

\bibitem[1995]{bwbl95}
Beauchamp, A., Wesemael, F., Bergeron, P., \& Liebert, J. 1995, \apj, 441, 85
\bibitem[1999]{bwbfs99}
Beauchamp, A., Wesemael, F., Bergeron, P., et al. 1999, \apj, 516, 887
\bibitem[1993]{Bradley93}
Bradley, P. A. 1993, Ph.D. thesis, University of Texas-Austin
\bibitem[1994]{bw94}
Bradley, P. A., \& Winget, D. E. 1994, \apj, 430, 850
\bibitem[1992]{bfwh94}
Brassard, P., Fontaine, G., Wesemael, F., \& Hansen, C. J. 1992, \apjs, 80, 369
\bibitem[1971]{bc17}
B\"{o}hm, K. H., \& Cassinelli, J. 1971, \aap, 12, 21
\bibitem[2005]{cnkwk05}
Castanheira, B. G., Nitta, A., Kepler, S. O., Winget, D. E., \& Koester, D. 2005, \aap, 432, 175 
\bibitem[2006]{ckhk06}
Castanheira, B. G., Kepler, S. O., Handler, G., \& Koester, D. 2006, \aap,  450, 331
\bibitem[2006]{elk06}
 Eisenstein, D. J., Liebert, J., Koester, D. 2006, \aj,  132, 676
\bibitem[1994]{gbdjms94}
Geist, A., Beguelin, A., Dongarra, J., et al. 1994, PVM: Parallel Virtual Machine, A User
Guide and Tutorial for Networked Parallel Computing (Cambridge: MIT Press)
\bibitem[2006]{jmgp06}
Gonzalez Perez, J. M., Solheim, J.-E., \& Kamben, R. 2006, \aap, 454, 527
\bibitem[2000]{hte00}
Han, Z.,  Tout, C. A., \& Egglenton, P. P. 2000, \mnras , 319, 215
\bibitem[2002]{hmw02} 
Handler, G., Metcalfe, T. S., \& Wood, M. A. 2002, \mnras, 335, 698 
\bibitem[1976]{i76}
 Iben, I. Jr. 1976, \apj , 208, 165
\bibitem[1982]{i82}
 Iben, I. Jr. 1982, \apj , 259, 244
\bibitem[1986]{i86}
 Iben, I. Jr. 1986, \aj , 304, 201 (I86)
\bibitem[1985]{it85}
Iben, I. Jr., Tutukov, A. V. 1985, \apjs, 58, 661 (IT85)
\bibitem[1996]{ir96}
Iglesias, C. A., \& Rogers, F. J. 1996, \apj, 464, 943
\bibitem[1996]{ihnk96} 
Itoh, N., Hayashi, H., Nishikawa, A., \& Kohyama, Y. 1996, \apjs, 102, 411
\bibitem[1984]{Kepler84}
 Kepler, S. O. 1984, \apj, 286, 314
\bibitem[1986]{Kawaler86} 
 Kawaler, S. 1986, Ph.D. thesis, University of Texas-Austin
 \bibitem[2003]{Kepler03}
Kepler, S. O., Nather, E. N., Winget, D. E., et al. 2003, \aap, 401, 639
 \bibitem[2007]{Kepler07}
Kepler, S. O., Kleinman, S. J., Nitta, A., et al. 2007, \mnras, 375, 1315 
\bibitem[2004]{khe04} 
Kleinman, S. J., Harris, H. C., Eisenstein, D. J., et al. 2004, \apj, 607, 426
\bibitem[1969]{ks69} 
 Kutter, G. S., \& Savedoff, M. P. 1969, \apj, 156, 1021
\bibitem[1975]{lvh75}
Lamb, D. Q., \& Van Horn, H. M. 1975, \apj, 200, 306
\bibitem[1986]{Liebert86} 
 Liebert J. 1986, Proc. IAU Coll. 87. D. Reidel Publishing Co., Dordrecht, p. 367
\bibitem[1999]{Metc99a} 
 Metcalfe, T. S., \& Nather, R. E. 1999, Linux J., 65, 58
\bibitem[1999]{Metc99b} 
Metcalfe, T. S., \& Nather, R. E. 2000, Baltic Astron., 9, 479
\bibitem[2000]{Metc00} 
 Metcalfe, T. S., Nather, R. E., \& Winget, D. E. 2000, \apj, 545, 974 (M0)
\bibitem[2001]{Metc01}  
 Metcalfe, T. S., Winget, D. E., \& Charbonneau, P. 2001, \apj, 557, 1021 (M1)
\bibitem[2005]{Metc05}  
 Metcalfe, T.~S., Nather, R.~E., Watson, T.~K., et al. 2005, \aap, 435, 649
\bibitem[1998]{Montgomery98}
 Montgomery, M. H. 1998, Ph.D. thesis, University of Texas-Austin
\bibitem[2001]{mmw01}
Montgomery M. H., Metcalfe T. S., \& Winget D. E. 2001, \apj, 548, L53
\bibitem[1981]{nather81}
Nather, R. E., Robinson, E. L., \& Stover, R. J. 1981, \apj, 244, 269
\bibitem[1990]{nather90}
 Nather, R. E., Winget, D. E., Clemens, J. C., Hansen, C. J., \& Hine, B. P. 1990, \apj, 361, 309
\bibitem[1997]{sdgbhim97} 
Salaris, M., Dominguez, I., Garcia-Berro, E., et al. 1997, \apj, 486, 413 
\bibitem[1997]{Shipman97}
Shipman, H. 1997, in White dwarfs: Proc. 10th European Workshop on White Dwarfs. Kluwer, Dordrecht, p.165 
\bibitem[1982]{rkn82} 
Robinson, E. L., Kepler, S. O., \& Nather, R. E. 1982, \apj, 259, 219
\bibitem[1995]{sch95} 
Saumon, D., Chabrier, G., \& van Horn, H. M. 1995, \apjs, 454, 527
\bibitem[2000]{Vuille00} 
Vuille, F., O\textquoteright Donoghue, D., Buckley, D. A. H., et al., 2000, \mnras, 314, 689
\bibitem[1981]{Winget81}
Winget, D. E. 1981, Ph.D. thesis, University of Rochester
\bibitem[1985]{wrnkd85}   
Winget, D. E., Robinson, E. L., Nather, R. E., et al. 1985, \apj, 292, 606
\bibitem[1994]{Winget94}   
Winget D. E., et al. 1994, \apj, 430, 839
\bibitem[2002]{wkmw02}
Wolff, B., Koester, D., Montgomery, M. H., \& Winget, D. E. 2002, \aap, 388, 320
\bibitem[1990]{Wood90}
Wood, M. 1990, Ph.D. thesis, University of Texas-Austin




\end{thebibliography}
\end{document}